\documentclass[twocolumn]{aastex6}

\slugcomment{Accepted for publication in The Astrophysical Journal}

\shorttitle{High neutronization in Supernova Remnants}
\shortauthors{H. Mart\'{i}nez-Rodr\'{i}guez et al.}

\usepackage{graphicx}	
\usepackage{amsmath}	
\usepackage{amssymb}	
\usepackage{xcolor}
\PassOptionsToPackage{hyphens}{url}\usepackage{hyperref}  
\hypersetup{
	colorlinks   = true, 
	urlcolor     = blue, 
	linkcolor    = blue, 
	citecolor   = blue 
}

\usepackage{float}
\usepackage{footmisc} 
\usepackage{verbatim}

\newcommand{\mch}{M$_{\rm Ch}$}

\begin{document}


\title{Observational evidence for high neutronization in supernova remnants: implications for Type Ia supernova progenitors}


\author{H\'{e}ctor Mart\'{i}nez-Rodr\'{i}guez\altaffilmark{1}, Carles Badenes\altaffilmark{1}, Hiroya Yamaguchi\altaffilmark{2,3}, Eduardo Bravo\altaffilmark{4},
F. X. Timmes\altaffilmark{5,6}, Broxton J. Miles\altaffilmark{7}, Dean M. Townsley\altaffilmark{7}, Anthony L. Piro\altaffilmark{8}, Hideyuki Mori\altaffilmark{9,10}, 
Brett Andrews\altaffilmark{1}, and Sangwook Park\altaffilmark{11}}



\altaffiltext{1}{Department of Physics and Astronomy and Pittsburgh Particle Physics, Astrophysics and Cosmology Center (PITT PACC), University of Pittsburgh, 3941 O'Hara Street, Pittsburgh, PA 15260, USA, \href{mailto:hector.mr@pitt.edu}{hector.mr@pitt.edu}}
\altaffiltext{2}{NASA Goddard Space Flight Center, Code 662, Greenbelt, MD 20771, USA}
\altaffiltext{3}{Department of Astronomy, University of Maryland, College Park, MD 20742, USA}
\altaffiltext{4}{E.T.S. Arquitectura del Vall\`{e}s, Universitat Polit\`{e}cnica de Catalunya, Carrer Pere Serra 1-15, E-08173 Sant Cugat del Vall\`{e}s, Spain}
\altaffiltext{5}{The Joint Institute for Nuclear Astrophysics, USA}
\altaffiltext{6}{School of Earth and Space Exploration, Arizona State University, Tempe, AZ, USA}
\altaffiltext{7}{Department of Physics \& Astronomy, University of Alabama, Tuscaloosa, AL, USA}
\altaffiltext{8}{Carnegie Observatories, 813 Santa Barbara Street, Pasadena, CA 91101, USA}
\altaffiltext{9}{CRESST and X-ray Astrophysics Laboratory, NASA Goddard Space Flight Center, Code 602, Greenbelt, MD 20771, USA}
\altaffiltext{10}{Department of Physics, University of Maryland, Baltimore County, 1000 Hilltop Circle, Baltimore, MD 21250, USA}
\altaffiltext{11}{Department of Physics, University of Texas at Arlington, Box 19059, Arlington, TX 76019, USA}


\begin{abstract}
  The physical process whereby a carbon--oxygen white dwarf explodes as a Type Ia supernova (SN Ia) remains highly uncertain. The
  degree of neutronization in SN Ia ejecta holds clues to this process because it depends on the mass and the metallicity 
  of the stellar progenitor, and on the thermodynamic history prior to the explosion. We report on a new method to determine ejecta 
  neutronization using Ca and S lines in the X-ray spectra of Type Ia supernova remnants (SNRs). 
  Applying this method to \textit{Suzaku} data of Tycho, Kepler, 3C 397 and G337.2$-$0.7 in the 
  Milky Way, and N103B in the Large Magellanic Cloud, we find that
  the neutronization of the ejecta in N103B is comparable to that of Tycho and Kepler, which suggests that progenitor metallicity
  is not the only source of neutronization in SNe Ia. We then use a grid of SN Ia explosion models to infer the metallicities of 
  the stellar progenitors of our SNRs. The implied metallicities of 3C 397, G337.2$-$0.7, and N103B are major outliers compared to 
  the local stellar metallicity distribution functions, indicating that progenitor metallicity can be ruled out as the origin 
  of neutronization for these SNRs. Although the relationship between ejecta neutronization and equivalent progenitor metallicity is 
  subject to uncertainties stemming from the $^{12}$C$\,$+$^{16}$O reaction rate, which affects the Ca/S mass ratio, our main results 
  are not sensitive to these details.
\end{abstract}





\section{Introduction} \label{Introduction}

\setcounter{footnote}{10} 

Type Ia supernovae (SNe Ia) are the thermonuclear explosions of white dwarf (WD) stars that are destabilized by mass accretion
from a close binary companion. Despite their importance for many fields of astrophysics, such as galactic chemical evolution 
\citep{Ko06, An16}, studies of dark energy \citep{Ri98,Pe99} and constraints on $\Lambda$CDM parameters \citep{Be14,Re14}, 
key aspects of SNe Ia remain uncertain, including the precise 
identity of their stellar progenitors and the mechanism that triggers the thermonuclear runaway. Discussions of SN Ia progenitors 
are often framed by the single degenerate and double degenerate scenarios, depending on whether the WD companion is a non-degenerate 
star or another WD. In the single degenerate scenario, the WD grows in mass through accretion over a relatively long timescale (t$ \, {\sim} \, 10^{6}$ year) 
and explodes when it gets close to the Chandrasekhar limit M$_{\rm{Ch}} \, {\simeq} \, 1.4 \, M_{\odot}$ \citep{No84,Th86,Ha96,Han04}. 
In most double degenerate scenarios, by contrast, the destabilizing event (a merger or collision) happens on a dynamical timescale 
\citep{Ib84}, quickly leading to an explosion that is not necessarily close to \mch $\,$ \citep{Ras09, Ross09, Ras10, Si10,vaK10, Ku13}. In principle, it is 
possible to discriminate between single degenerate and double degenerate systems exploding on a dynamical timescale after merging, provided that some 
observational probes are sensitive to the presence or absence of an extended accretion phase leading to the thermonuclear runaway and to the mass of the 
exploding star (see the recent reviews by \citealt{Wa12} and \citealt{Ma14}). Here we examine one of these probes, the degree
of neutronization in SN Ia ejecta.

The neutron excess, defined as $\eta = 1 - 2Y_{\rm{e}} = 1 - 2\left<Z_{A}\right> / \left<A\right>$ (where $Y_{\rm{e}}$ is the electron fraction,
 $Z_{A}$ is the atomic number, and $A$ is the mass number) 
should be zero in WDs composed solely of $^{12}$C and $^{16}$O. The value of $\eta$ can be increased through weak interactions 
taking place at different stages during the life of SN Ia progenitors. So far, three such mechanisms have been proposed.

\placefigure{f1}
\begin{figure*}
\centering
\includegraphics[scale = 0.30]{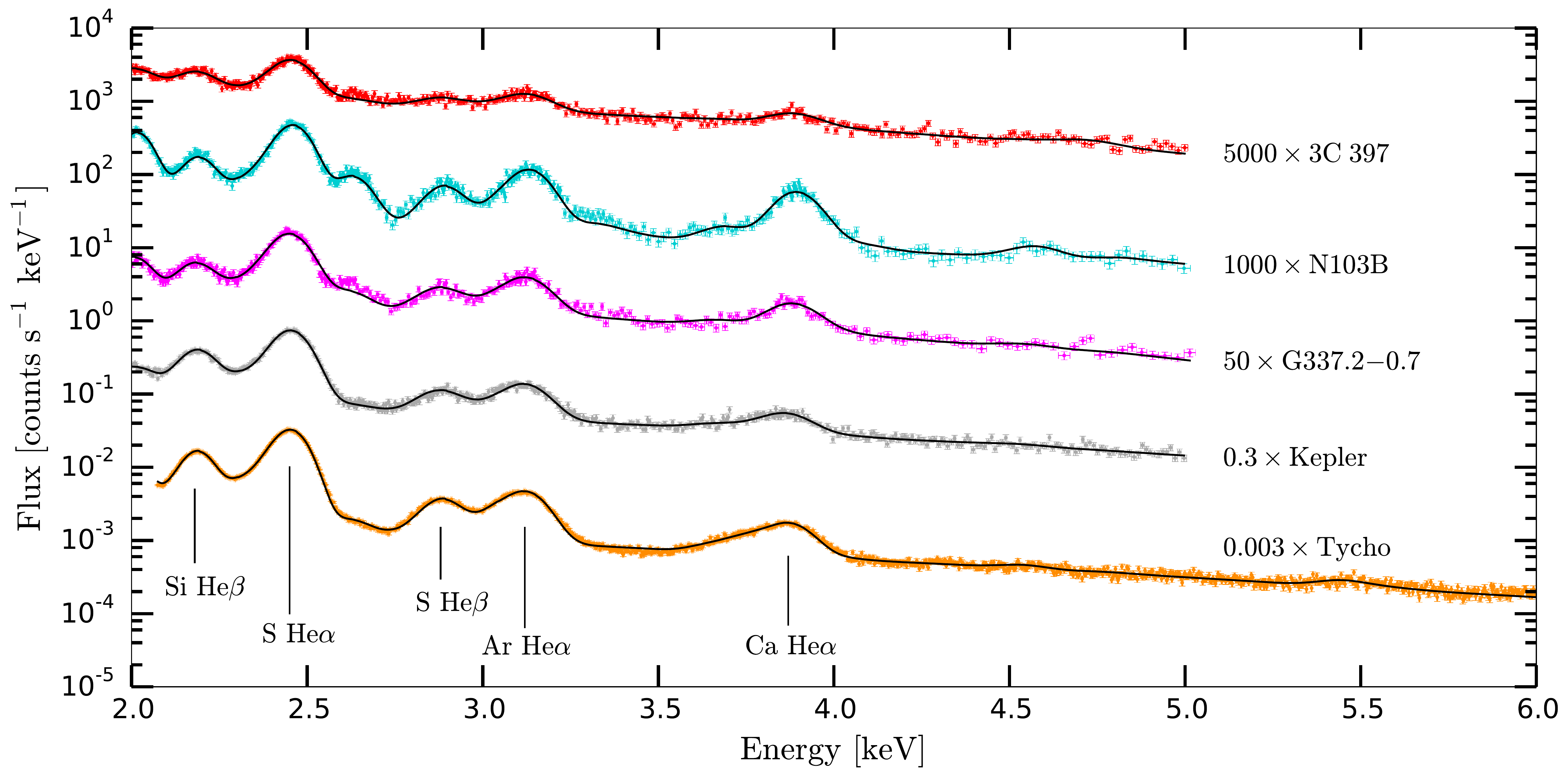}
\caption{\textit{Suzaku} XIS0 and XIS3 combined spectra of 3C 397, N103B, G337.2$-$0.7, Kepler and Tycho between 2.0 and 5.0
  keV. The SNRs are sorted in decreasing order of Fe ionization state \citep{Ya14}. The most relevant atomic transitions are
  labeled. For Tycho, it is necessary to extend the upper energy limit from 5.0 to 6.0 keV in order to achieve a reduced 
  chi-square $\chi^{2}/\nu < 2$.}
\label{Spectrum}
\end{figure*}

\begin{enumerate}

\item{\textbf{Progenitor metallicity}}. The bottleneck reaction in the CNO cycle, $^{14}$N(p,$\gamma$)$^{15}$O, causes all the
  C, N, and O in the progenitor to pile up onto $^{14}$N at the end of H burning, which then becomes $^{22}$Ne during hydrostatic He
  burning through the chain $^{14}$N($\alpha$, $\gamma$)$^{18}$F($\beta^{+}$, $\nu_{e}$)$^{18}$O($\alpha$, $\gamma$)$^{22}$Ne. Since
  $^{22}$Ne carries a neutron excess, this results in a linear scaling of $\eta$ with progenitor metallicity $Z$: $\eta = 0.1 Z$
  \citep{Ti03,Br10, MoRa16}. Hence, this $^{22}$Ne content is usually defined as the ``metallicity'' of a WD.

\item{\textbf{Carbon simmering}}. In SN Ia progenitors that approach \mch\ through slow accretion, carbon can ignite close to the center without
  immediately triggering a thermonuclear runaway. Instead, the WD develops a large ($ {\sim} \, 1 \, M_{\odot}$) convective core for 
  a few thousands of years until the heat from fusion overwhelms neutrino cooling and an explosion ensues \citep{Wo04,Wu04,PiC08}. During this `C simmering' phase, 
  electron captures on the products of C fusion (mostly $^{13}$N and $^{23}$Na) increase the value of $\eta$ \citep{Ch08,PiB08,MR16}.

\item{\textbf{Neutron-rich Nuclear Statistical Equilibrium}} (n-NSE). When a WD explodes close to \mch, the inner ${\sim} \, 0.2 \, M_{\odot}$
  is dense enough for electron captures to take place during nucleosynthesis, shifting the equilibrium point of NSE away from
  $^{56}$Ni to more neutron-rich species like $^{55}$Mn and $^{58}$Ni \citep{Iw99,Br00}.

\end{enumerate}

To summarize, the baseline neutronization level in all SNe Ia is set by progenitor metallicity. Additional neutronization can be
introduced only in systems that explode close to \mch, by C simmering or n-NSE. C simmering will affect most of the SN ejecta,
while n-NSE will only affect the NSE material synthesized in the innermost layers (i.e., Fe-peak elements). Thus, while mixing 
may blur this distinction to some degree, accurate measurements of $\eta$ in SNe have the potential to constrain the fundamental 
properties of SN Ia progenitors.

Emission lines from stable Mn and Ni in the X-ray spectra of Type Ia supernova remnants (SNRs) have been used to
measure $\eta$ and infer the properties of SN Ia progenitors \citep{Ba08,Park13,Ya15}. However, these weak lines are often hard to
detect, and it is difficult to disentangle the neutronization effects of n-NSE and C simmering using Fe-peak nuclei \citep[see][for
discussions]{Park13,Ya15}. Here, we report on a new method to measure neutronization in SNe Ia based on the sensitivity of the Ca/S yield to $\eta$ identified 
by \cite{De14}. $\rm{^{40}Ca}$ and $\rm{^{32}S}$ are produced in a quasi-nuclear statistical equilibrium in a temperature range $\simeq$ 2$-$4$\times$10$^{9}$ K. 
	In this regime, the nuclear abundances are determined by a set of coupled Saha equations that ultimately depend on the temperature, density, and Y$_e$ 
	\citep[e.g.][]{Cl65,Har85,Na04,Se08,De14}. Thus, the abundances of symmetric nuclei such as $^{32}$S and $^{40}$Ca depend on the overall Y$_e$. 
	For explosive events such as SNe Ia, the freeze-out from high temperatures occurs on a time scale faster than the nuclear rearrangement, ensuring that the 
	abundances produced at these temperatures are the same as the final abundances \citep{De14,Mi16}. Among the intermediate-mass elements, $\rm{^{40}Ca}$ is 
the most sensitive to changes in the electron fraction. \cite{De14} found a systematic quasi-linear 
 $\rm{^{32}S}$ yield with respect to $Y_{\rm{e}}$, and a 
more complex trend for the global abundance of $\rm{^{40}Ca}$. Thus, more neutron-rich progenitors should have a lower Ca/S mass ratio ($\rm{M_{Ca}/M_{S}}$). 
Here we show that the Ca/S mass ratio in SN Ia ejecta is indeed a good observational tracer of neutronization, 
with the key advantages that (a) it is not affected by n-NSE and (b) it uses much stronger 
emission lines that can be easily measured in a larger sample of objects.

This paper is organized as follows. In Section \ref{Observational}, we describe the observations and derive  $\rm{M_{Ca}/M_{S}}$ values. 
In Section \ref{Interpretation}, we interpret the inferred $\rm{M_{Ca}/M_{S}}$ and discuss the implications for SN Ia physics. In Section \ref{Rate},
we analyze the relation between $\rm{M_{Ca}/M_{S}}$ and the $^{12}$C$\,$+$^{16}$O reaction rate.
Finally, in Section \ref{Conclusions}, we summarize our results and outline future lines of work.

\section{Observations and data analysis} \label{Observational}

\cite{Ya14} list 11 Type Ia SNRs with Fe K$\alpha$ emission in the Milky Way and the Large Magellanic Cloud (LMC). We
re-reduce and analyze all these \textit{Suzaku} spectra, paying special attention to the emission lines from S, Ar, Ca, Cr, Mn,
Fe, and Ni. We do not include Si in our analysis because of the well-known calibration problems around ${\sim} \, 1.5$ keV in the
\textit{Suzaku} CCDs (\url{https://heasarc.gsfc.nasa.gov/docs/suzaku/analysis/sical.html}).

\begin{table*}
\begin{center}
  \caption{Summary of the \textit{Suzaku} spectral modeling for the SNRs shown in Figure \ref{Spectrum}. 
  	See Table 1 from \cite{Ya14} for a list of the observation IDs and dates corresponding to each SNR. \label{Measurements}}
\begin{tabular}{cccccccc}
\tableline\tableline
\noalign{\smallskip}
SNR & Exp. Time & $N_{\rm{H}}$ & Continuum Model & Refs.\footnote{References consulted for the absorption and continuum components in the spectral fittings: (1) \cite{Sa05}, (2) \cite{Le03}, (3) \cite{So14}, (4) \cite{Ra06}, (5) \cite{Rey07}, (6) \cite{Pat12}, (7) \cite{Park13}, (8) \cite{Ba06}.} & $\rm{M_{Ar}/M_{S}}$\footnote{\label{refnote}All the uncertainties are in the 90\% confidence range ($\Delta \chi^{2} = 2.706$). Note that the confidence intervals do not necessarily have to be symmetric (\url{https://heasarc.gsfc.nasa.gov/xanadu/xspec/manual/XSerror.html}).} & $\rm{M_{Ca}/M_{S}}$\footref{refnote} & $\rm{M_{Cr}/M_{Fe}}$\footref{refnote} \\
\noalign{\smallskip}
 & (ks) & $(10^{22} \, \rm{cm}^{-2})$ & & & & & \\
\noalign{\smallskip}
\tableline
\noalign{\smallskip}
3C 397 & 104 & 3.00 & Bremms. ($kT = 0.16 $ keV\footnote{Best-fit parameter.}) & 1 & $0.214^{+0.030}_{-0.026}$ & $0.213^{+0.021}_{-0.034}$ & $0.040^{+0.029}_{-0.016}$ \\
\noalign{\smallskip}
N103B & 224 & 0.34 & Power law ($\Gamma = 3.70$) & 2,3 & $0.257^{+0.024}_{-0.035}$ & $0.255^{+0.021}_{-0.036}$ & $0.028^{+0.021}_{-0.014}$ \\
\noalign{\smallskip}
G337.2$-$0.7 & 304 & 3.20 & Power law ($\Gamma = 2.20$) & 4 & $0.214^{+0.016}_{-0.013}$ & $0.169^{+0.016}_{-0.023}$ & Undeterm. \\
\noalign{\smallskip}
Kepler & 146 & 0.52 & Power law ($\Gamma = 2.67$) & 5,6,7 & $0.279^{+0.010}_{-0.017}$ & $0.283^{+0.016}_{-0.023}$ & $0.008^{+0.007}_{-0.005}$ \\
\noalign{\smallskip}
Tycho & 313 & 0.60 & Power law ($\Gamma = 2.54$) & 8 & $0.218^{+0.022}_{-0.010}$ & $0.252^{+0.025}_{-0.011}$ & $0.016^{+0.018}_{-0.005}$ \\
\noalign{\smallskip}
\tableline
\end{tabular}
\end{center}
\end{table*}

We merge the data from the two active front-illuminated CCDs (XIS0 and 3) to increase photon counts. The spectrum of each SNR is
fit in the 2.0$-$5.0 keV energy range with a plane-parallel shock model 
\citep[vvpshock,][]{Bo01} plus an additional component for the
continuum (either bremsstrahlung or a power law), using the \texttt{XSPEC} software \citep[][version 12.9.0i, \url{https://heasarc.gsfc.nasa.gov/xanadu/xspec/manual/}]{Ar96} 
and the most recent non-equilibrium ionization atomic data from \textit{AtomDB} \citep{Fo12, Fo14}. We fix the hydrogen column densities $N_{\rm{H}}$ and the continuum components 
to values previously reported for each SNR (see references in Table \ref{Measurements}). We let the electron temperature $T_{e}$, the ionization time scale $n_{e}t$
(defined as the product of the electron density and the expansion age) 
and the abundances of the $\alpha$-elements in the shock model vary until we get a valid fit, with a reduced chi-square $\chi^{2}/\nu < 2$ (where $\nu$ is the number of degrees of freedom). 
This allows us to derive confidence intervals for the different parameters.
We convert these abundances retrieved by the  best-fit spectral model into mass ratios using the \cite{AG89} factors.

Our goal is to measure Ca/S mass ratios to better than ${\sim} \, 20\%$ in order to 
compare with a grid of SN explosion models (where physically meaningful variations of $\rm{M_{Ca}/M_{S}}$ are of this order or larger). 
Only the five objects shown in Figure \ref{Spectrum} pass this quality cut: 3C 397, G337.2$-$0.7, Kepler and 
Tycho in the Milky Way, and N103B in the LMC. The relevant parameters for the observations are listed in Table \ref{Measurements}. Additionally, we determine Ar/S mass ratios for these SNRs.

As a sanity check, we also fit all spectra using two single-ionization timescale non-equilibrium ionization models  
\citep[vvrnei, ][]{Hu00}, and find mass
ratios consistent with the values obtained with the plane-parallel shock models. For Tycho, we are unable to get a valid fit with a plane-parallel shock model, 
so we use two non-equilibrium ionization models in an enlarged energy window between $2.0$ and $6.0$ keV (See Figure \ref{Spectrum}). Only this spectral model can successfully fit 
the Ca He$\alpha$ feature \citep[see][for a discussion about this line in the spectrum of Tycho and the difficulties to reproduce it with explosion models]{Ba06}. 
We follow the same procedure around the Fe K$\alpha$ line (5.0$-$8.0 keV) for each SNR, but can confidently detect the Mn and Ni lines only 
for 3C 397, Kepler and Tycho \citep[measurements reported in][]{Ya15}, so we choose to determine the Cr/Fe mass ratio ($\rm{M_{Cr}/M_{Fe}}$) for all objects as a baseline measurement of Fe-peak ejecta.

The final Ar/S, Ca/S and Cr/Fe mass ratios are listed in Table \ref{Measurements}. The relative errors in the inferred $\rm{M_{Ca}/M_{S}}$ are in the range of ${\sim} \, $5$-$16\%, which
allows for meaningful comparisons with explosion models. These are lower than the previous Fe-peak relative errors for 3C 397, Kepler and Tycho: ${\sim} \, $35$-$70\% 
($\rm{M_{Mn}/M_{Cr}}$) and ${\sim} \, $28$-$65\% ($\rm{M_{Mn}/M_{Fe}}$, $\rm{M_{Ni}/M_{Fe}}$). Hence, the mass ratios of intermediate-mass elements can be measured with better 
precision than those of Fe-peak elements. Prior measurements of $\rm{M_{Ca}/M_{S}}$ in the optical spectra of SNe Ia are based on tomography (e.g. 0.299 for SN 2002bo, \citealt{Ste05}; 0.029 for
SN 2003du, \citealt{Ta11}; between $ \, 0.250^{+0.088}_{-0.088} \, $ and $ \, 0.40^{+0.14}_{-0.14} \, $ for SN 1986G, \citealt{Ash16}), and strongly depend on the radiative transfer treatment and
on the chosen explosion model. The error bars from these tomography estimates are either undetermined or higher (${\sim} \, $35\%) than our measured errors. It is worth mentioning that all these 
measurements, with the exception of SN 2003du, overlap the ones reported in this paper (see Figure \ref{Comparison}).

\placefigure{f2}
\begin{figure*}
\centering
\hspace{-0.75 cm}
\includegraphics[scale=0.60]{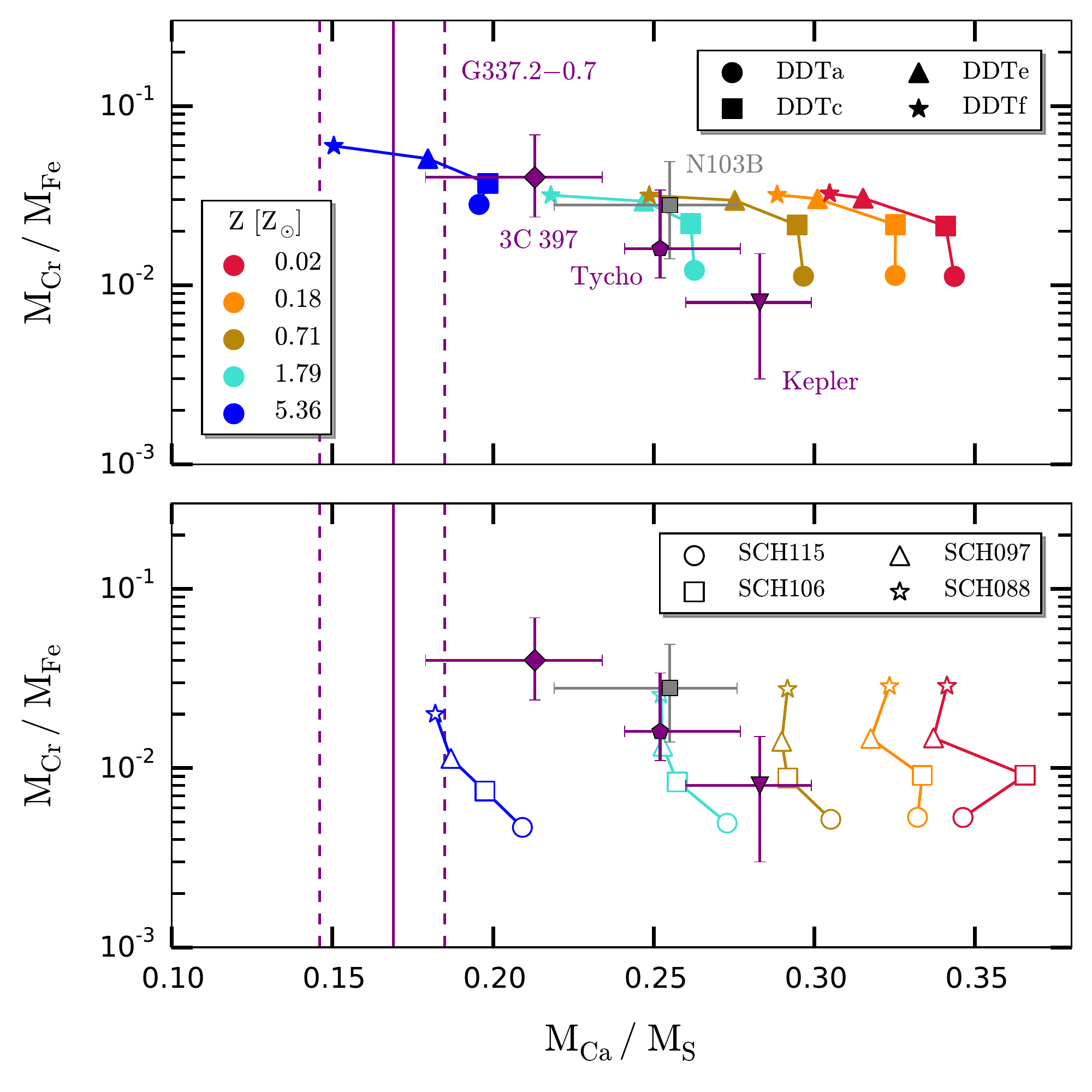}
\caption{$\rm{M_{Cr}/M_{Fe}}$ vs. $\rm{M_{Ca}/M_{S}}$ for 3C 397, N103B, Kepler and Tycho (Table \ref{Measurements}), compared with the 
theoretical predictions from SN Ia models (see Section \ref{Interpr_models}). The purple, vertical lines correspond to $\rm{M_{Ca}/M_{S}}$ for G337.2$-$0.7, whose $\rm{M_{Cr}/M_{Fe}}$ 
could not be determined. Top: \mch\ models. Bottom: sub-\mch\ models.}
\label{CrFeCaS}
\end{figure*}

Before doing a direct comparison between models and SNR observations, we must distinguish between dynamically old objects like 3C 397
and G337.2$-$0.7, which have likely thermalized the entire SN ejecta \citep{Ra06,Ya15}, and dynamically young objects like Kepler
and Tycho, which probably have not \citep{Ba06,Pat12}, with N103B being a transitional object between the two classes
\citep{Le03,Wi14}. The X-ray spectra of dynamically young objects are only representative of the shocked material, not of the entire
SN ejecta, and comparisons to bulk yields from SN explosion models should be done with some caution. However, the diagnostic Ca/S
mass ratios in Kepler and Tycho are largely unaffected by this, since the vast majority of the explosive Si-burning material
has already been shocked in these two objects \citep{Ba06,Pat12}. 

The $\rm{M_{Ca}/M_{S}}$ values measured in our SNRs span the range between 0.17 and 0.28. N103B has 
$\rm{M_{Ca}/M_{S}} \approx \ $0.26, between Tycho (0.25) and Kepler (0.28). This alone makes it challenging to invoke progenitor 
metallicity as the only source of neutronization in SN ejecta \citep[e.g.,][]{Ti03}, unless Kepler's progenitor was
 more metal-poor than most LMC stars, which seems unlikely given its measured Fe-peak mass ratios \citep{Park13} and location toward
 the Galactic center region. Therefore, our observations alone, without any comparison to models, indicate that progenitor
 metallicity is not the only source of neutronization in SN Ia progenitors.

\placefigure{f3}
\begin{figure}
\centering
\hspace{-1 cm}
\includegraphics[scale=0.50]{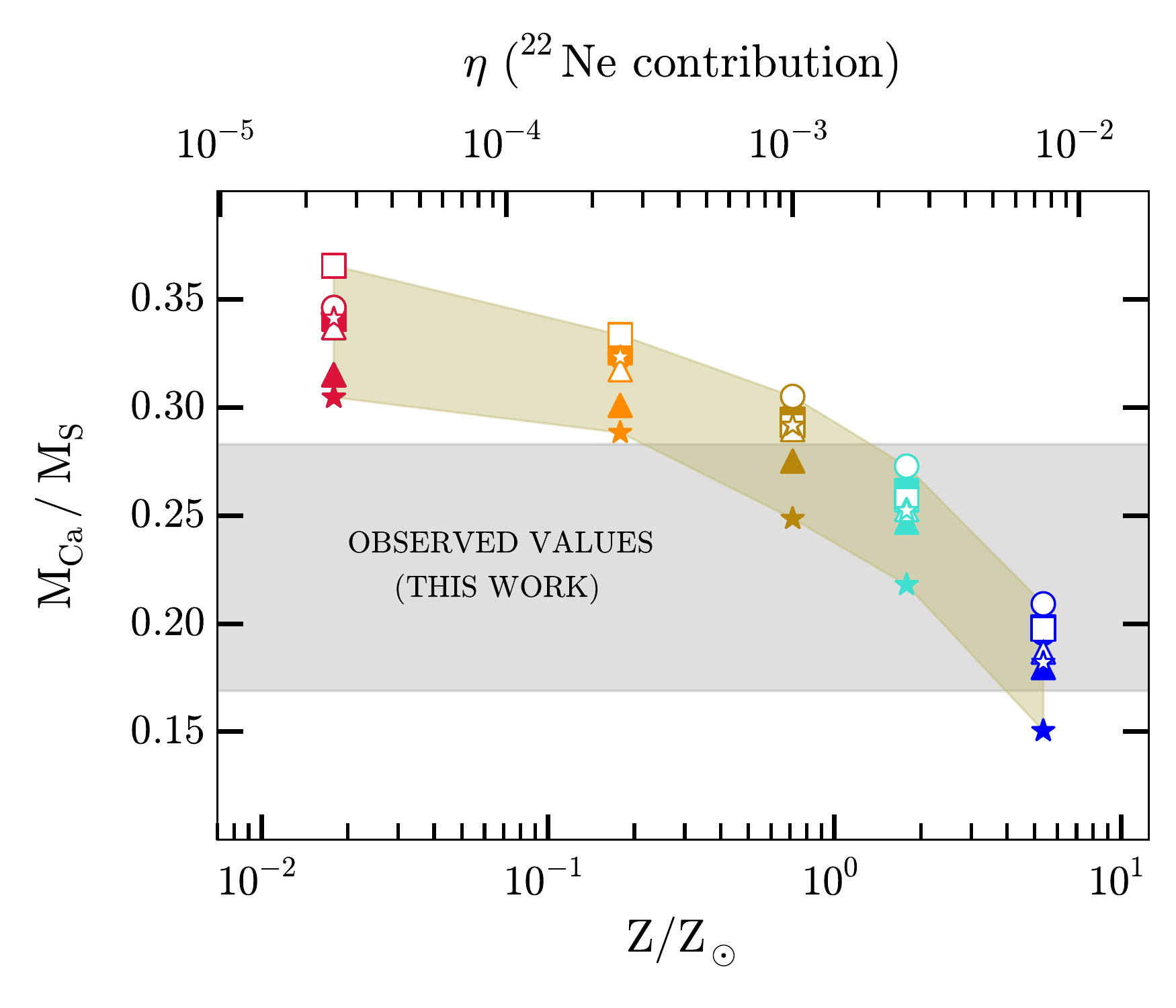}
\caption{$\rm{M_{Ca}/M_{S}}$ vs. progenitor metallicity for the models depicted in Figure \ref{CrFeCaS}. 
		Our measured mass ratios are shown as a gray, shaded strip, and the khaki region covers the theoretical predictions
		from the models. The neutron excess $\eta$ is given above the panel. Here, $\eta = 0.1 Z$,
		showing the $\rm{^{22}Ne}$ contribution to the overall neutronization \citep{Ti03}, because our models do not include the effect of
		C simmering (Section \ref{Interpr_models}) and $\rm{M_{Ca}/M_{S}}$ is not affected by n-NSE (Section \ref{Introduction}). More neutron-rich
		progenitors have a lower $\rm{M_{Ca}/M_{S}}$.}
\label{CaSeta}
 \end{figure}

\section{Interpretation} \label{Interpretation}

\subsection{Comparison with explosion models} \label{Interpr_models}

To interpret our measured mass ratios, we use the spherically symmetric SN Ia explosion models introduced in
\citet{Ya15}, which are calculated with a version of the code described in \citet{Br12}, updated to account for an accurate
coupling between hydrodynamics and nuclear reactions \citep{Br16}. In this model grid, the \mch\ explosions are 
delayed detonations \citep{Kh91} with a central density $\rho_{\rm{c}} = 2 \times 10^{9} \, \rm{g \, cm^{-3} }$ 
	and different deflagration-to-detonation densities ($\rho_{\rm{DDT}}$): $3.9, 2.6, 1.3$ and $1.0
\times 10^{7} \, \rm{g \, cm^{-3}}$, labeled as DDTa, DDTc, DDTe, and DDTf \citep[see][]{Ba03, Ba05, Ba08b}. The sub-\mch\ models are
central detonations of CO WDs with a core temperature $T_{\rm{c}} = 10^{8} \, \rm{K}$ and masses 
$ M_{\rm{WD}} = 0.88, 0.97, 1.06$  and $1.15 \, M_{\odot}$, similar to the models by
\cite{Si10}. Each model in the grid is calculated with five different values of the progenitor metallicity, $Z =
0.02, 0.18, 0.71, 1.8 $ and $ 5.4 \, Z_{\odot}$, taking $Z_{\odot} = 0.014$ \citep{As09}. This progenitor neutronization is
set by increasing the abundance of $^{22}$Ne in the pre-explosion WD according to the \cite{Ti03} metallicity relation. Additional neutronization
from C simmering in \mch\ models with large convective cores ($ {\sim} \, 1 \, M_{\odot}$) should behave in a similar way, i.e., 
increasing the value of $\eta$ throughout the convective region of the pre-explosion WD \citep{MR16}. However, for simplicity, we have not 
included a separate enhancement of $\eta$ due to simmering in this model grid. Because no simmering is included in our models, the
	level of neutronization in intermediate-mass elements is controlled exclusively by progenitor metallicity. The value of $\eta$ in the 
inner $ {\sim} \, 0.2 \, M_{\odot}$ of ejecta in the \mch\ models is further modified by n-NSE nucleosynthesis during the explosion 
\citep{Iw99, Br00}. Although simplified, this model grid captures the basic phenomenology of neutronization in SN Ia progenitors.

The bulk Cr/Fe vs. Ca/S mass ratios in the models are shown in Figure \ref{CrFeCaS}, together with the values measured in the five
SNRs in our sample. As expected from \cite{De14}, the Ca/S mass ratio in the models is a good tracer of progenitor 
neutronization (see also Figure \ref{CaSeta}). 
Models with different metallicities that burn Ca and S at similar temperatures have $\rm{M_{Ca}/M_{S}}$ values that 
can be discriminated by observations. This is because the main contribution to both elements comes from the isotopes $\rm{^{40}Ca}$ 
and $\rm{^{32}S}$, whose abundances are in quasi-statistical equilibrium at the temperatures ($\simeq$ 4$\times$10$^{9}$ K) at which
 $\rm{^{40}Ca}$ is synthesized. In this regime, $\rm{M_{Ca}/M_{S}} \propto X_\alpha^2$, where $X_\alpha$ is the abundance of alpha 
 particles, which decreases as metallicity increases \citep[see Figures 2 and 8 in][]{Br13}. The DDT models with the lowest 
 $\rho_{\rm{DDT}}$ (DDTe and f), which correspond to the low luminosity end of SNe Ia, show lower Ca/S mass ratios 
 because they burn a larger mass of Ca at a lower density and temperature than their more energetic counterparts, 
 which results in a lower Ca/S mass ratio. Figure \ref{CaSeta} shows that, for a given metallicity, 
 	the \mch\ and sub-\mch\ models predict similar $\rm{M_{Ca}/M_{S}}$ values.

\placefigure{f4}
\begin{figure*}
\centering
\includegraphics[scale=0.60]{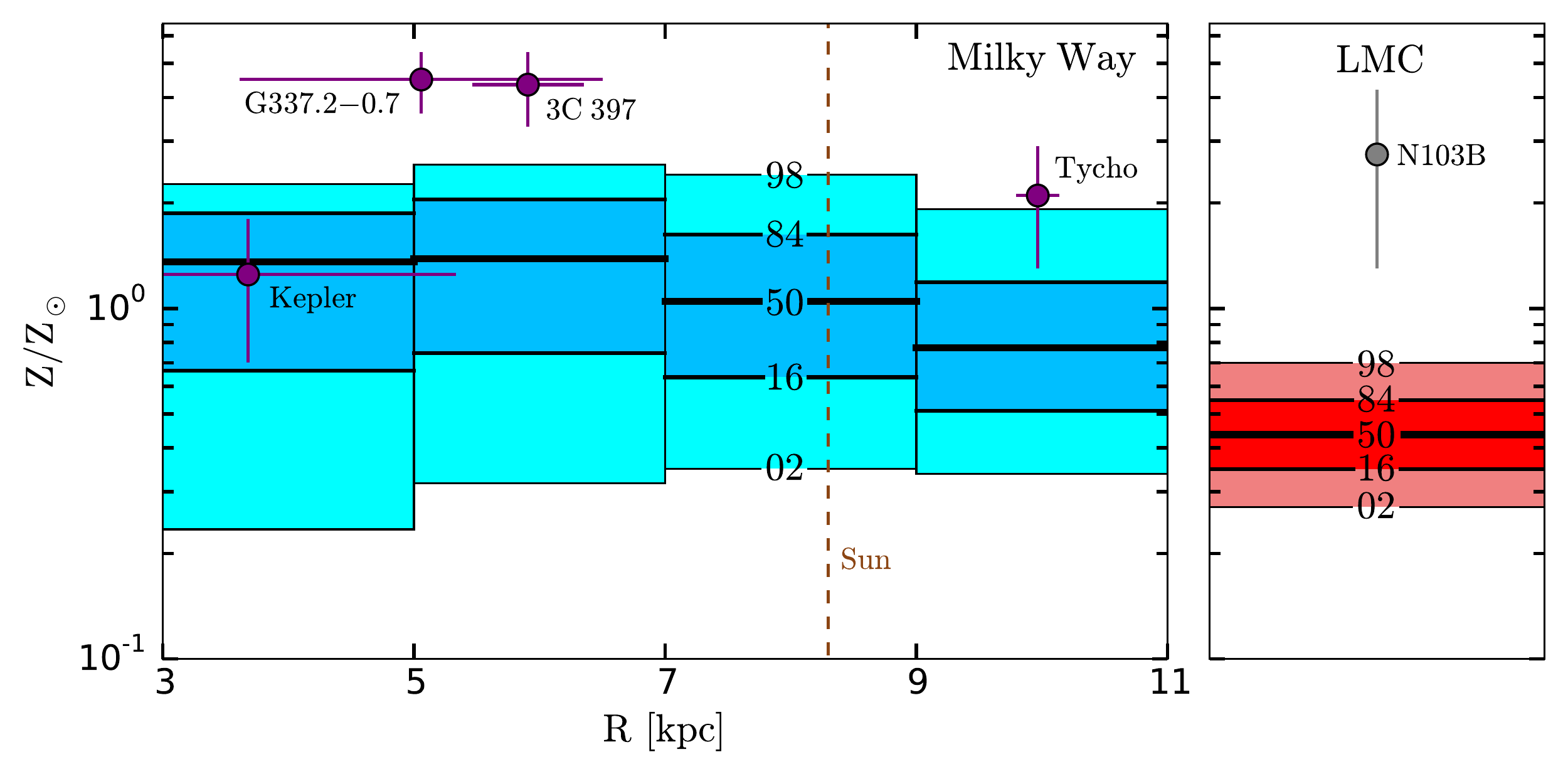}
\caption{Comparison between the implied metallicities of the SNRs and the stellar metallicity distributions (numbers indicate percentiles) 
	for the Milky Way (as a function of Galactocentric radius) and LMC disks. We consider a maximum height over the Milky Way disk $|z| = 0.6$ kpc, which
	encompasses the four Galactic SNRs. The solar Galactocentric distance \citep[8.3 kpc;][]{Gi09} is shown as a dashed, brown line.}
\label{MDF}
\end{figure*}

It is worth noting that the models in our grid span the observed Ca/S and Cr/Fe mass ratios for all the
SNRs. Furthermore, the level of neutronization inferred from the closest equivalent progenitor metallicity ($Z_{eq}$) is rather 
high in all SNRs. When compared to the metallicity distribution functions (MDFs) in the Milky Way and the LMC 
	(see Figure \ref{MDF} and Section \ref{Interpr_MDFs}), this suggests an additional
	source of neutronization in SN Ia ejecta. One possibility is carbon simmering. 
To quantify the increase in $Z_{eq}$, we need to consider some additional information 
about the objects under study. The properties of the Fe K$\alpha$ emission analyzed by \cite{Ya14} rule out the \mch\ models with the 
lowest $\rho_{\rm{DDT}} $ (DDTe and f) for N103B, 3C 397, Kepler, and Tycho, and favor them for G337.2$-$0.7. These constraints are 
confirmed by detailed spectral modeling for Tycho, G337.2$-$0.7 and Kepler \citep{Ba06,Ra06,Pat12}, and by the light echo spectrum 
of Tycho \citep{Kr08}. Once the ruled out \mch\ models are removed, we can better constrain the $Z_{eq}$ values for each SNR from the Ca/S mass ratio: 
$5.4 \, Z_{\odot}$ for 3C 397 and G337.2$-$0.7, $1.8 \, Z_{\odot}$ for Tycho and N103B, and between 1.8 and $0.7 \, Z_{\odot}$ for Kepler. 
These values are roughly the same for sub-\mch\ explosions, although the Cr/Fe mass ratio can rule out these models for 3C 397 
\citep[see also][]{Ya15}. We note that \cite{Vi16} proposed a sub-\mch\ progenitor for Kepler based on the properties of its light curve, 
and our measured $\rm{M_{Cr}/M_{Fe}}$ is in good accordance with the sub-\mch\ models in our grid. 
These $Z_{eq}$ results are in agreement with previous analyses based 
on emission lines from Fe-peak elements in Tycho, Kepler, and 3C 397 \citep{Ba08,Park13,Ya15}, but they represent a much cleaner measurement 
of the pre-explosion neutronization in the progenitor, since the Ca/S mass ratio is not susceptible to contamination from n-NSE material 
synthesized in the deepest layers of the WD \citep[see][ for a discussion]{Park13}.

\subsection{Comparison with metallicity distribution functions} \label{Interpr_MDFs}

The significance of the high values of $Z_{eq}$ that we infer from the X-ray spectra becomes apparent when we compare them to the
MDFs of the underlying stellar populations. This is shown in Figure \ref{MDF}, where we take
the MDF as a function of Galactocentric radius for the Milky Way disk \citep{Hay15} and the bulk MDF of the LMC \citep[adapted
from][]{Cho16}. The Galactocentric radii of the Milky Way SNRs  are calculated from their Galactic 
coordinates and the most recent estimates for their distances from the solar system: 6.5$-$9.5 kpc to 3C 397 \citep{Lea16}, 
2.0$-$9.3 kpc to G337.2$-$0.7 \citep{Ra06}, 3.0$-$6.4 kpc to Kepler \citep{Reyn99,San05} and 2.5$-$3.0 kpc to Tycho \citep{Tia11}.
We linearly interpolate between our DDT models (upper panel of Figure \ref{CrFeCaS}) to find an approximate $Z_{eq}$ range for each SNR,
excluding the models that can be ruled out based on the Fe K$\alpha$ emission. We note that the $Z_{eq}$ values
are similar in \mch\ and sub-\mch\ explosions (see Section \ref{Interpr_models}).

Our analysis indicates that progenitor metallicity can be ruled out as the only source of neutronization in 3C 397, G337.2$-$0.7, and N103B, 
which are many standard deviations above the mean stellar metallicity of their environments in the Galaxy and the LMC (see Figure \ref{MDF}). 
Tycho is a ${\sim} \ 2 \sigma$ outlier, and Kepler is the only object whose neutronization is compatible with the stellar metallicity distribution 
in its Galactic environment.

\section{Sensitivity of $\rm{M_{Ca}/M_{S}}$ to the {\rm $^{12}$C$\,$+$^{16}$O} reaction rate} \label{Rate}

\begin{figure*}
\centering
\includegraphics[scale=0.45]{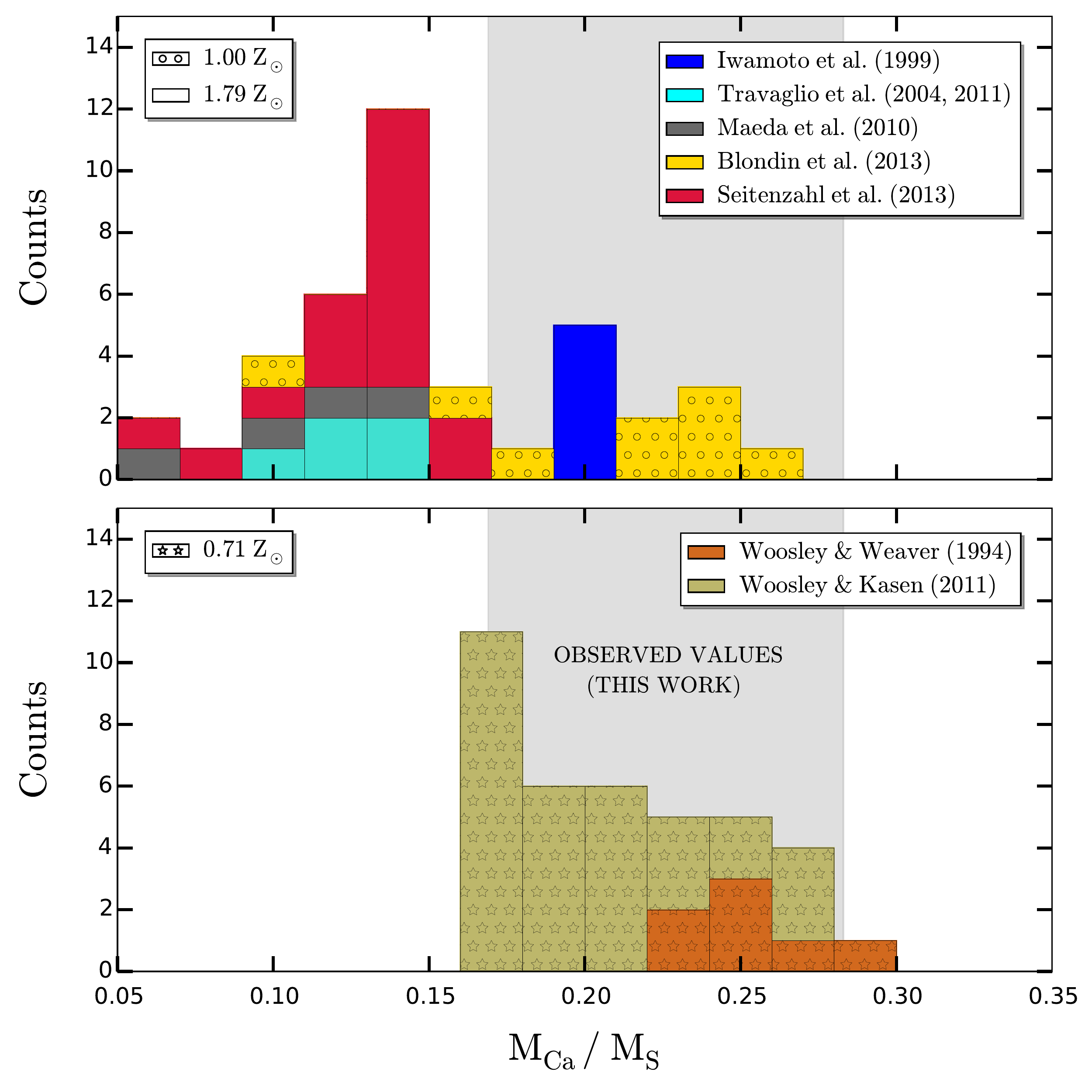}
\caption{Histogram for the Ca/S mass ratio predicted by various model grids from the literature. Top: \mch\ models. Bottom: sub-\mch\ models. 
Our measured values are depicted as a gray, shaded region.}
\label{Comparison}
\end{figure*}

Because a grid of SN Ia explosion models is needed to translate our
measured Ca/S mass ratios into equivalent progenitor metallicities, it
is important to verify the sensitivity of this ratio to the details of
explosive nucleosynthesis calculations. To this end, we compare our
observed $\rm{M_{Ca}/M_{S}}$ to the predictions of six
\mch\ \citep{Iw99, Tra04, Mae10, Tra11, Blo13, Se13} and two
sub-\mch\ \citep{Wo94, Wo11} SN Ia explosion model grids from the
literature.  Figure \ref{Comparison} shows that the multi-dimensional
models, \citet[2D and 3D]{Tra04}, \citet[2D]{Mae10}, \citet[2D]{Tra11}
and \citet[3D]{Se13} predict a Ca/S mass ratio that is substantially
(${\sim} \ 50 \%$) lower than both the models in our grid and the
observations, unlike the spherically symmetric calculations 
in 1D \citep{Wo94, Iw99, Wo11, Blo13}.

\placefigure{f5}

\placefigure{f6}
\begin{figure}
\centering
\hspace{-0.75 cm}
\includegraphics[scale=0.60]{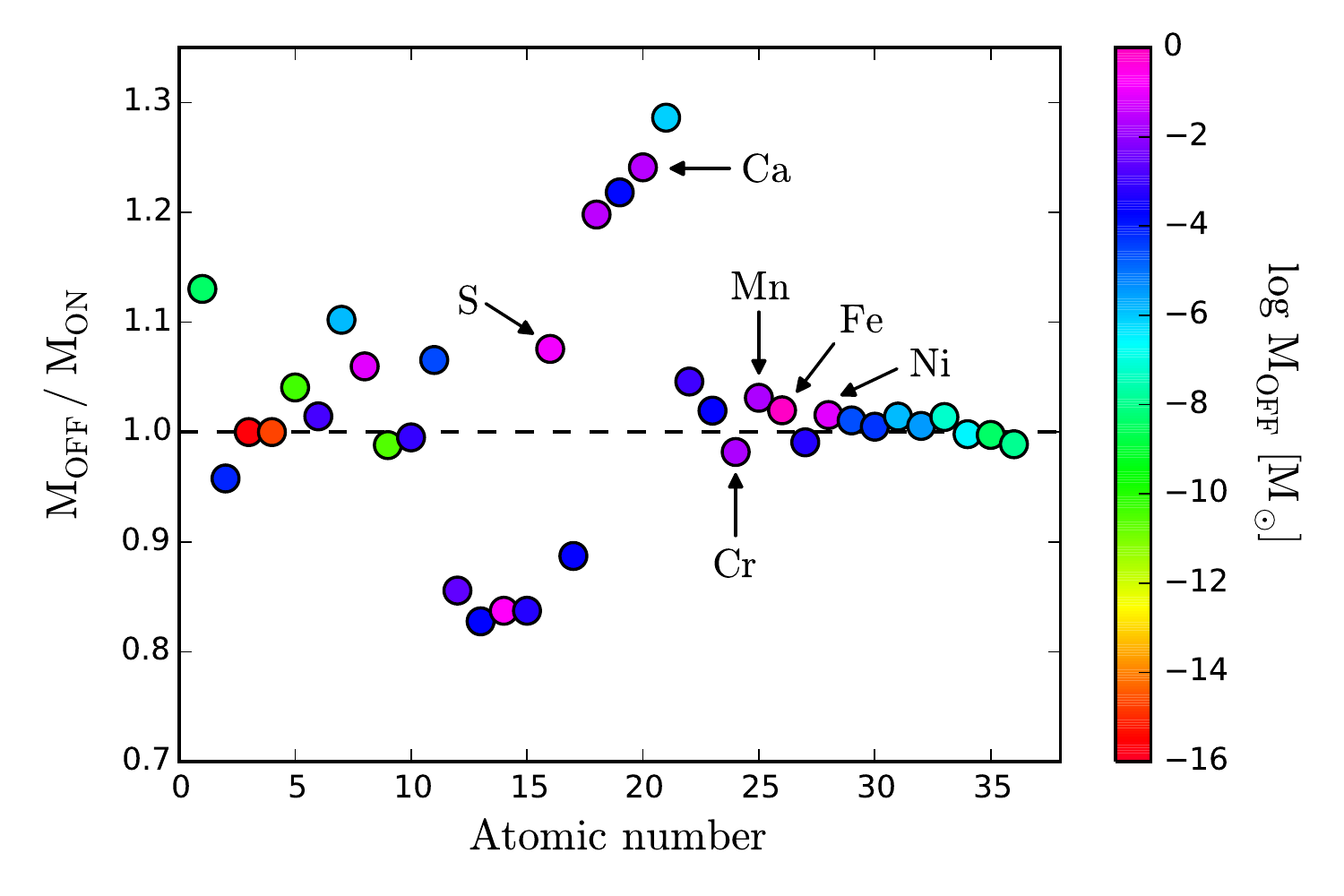}
\caption{Total yields spanning from hydrogen ($Z_{A} = 1$) to krypton ($Z_{A} = 36$) for two DDTc, 5.4-Z$_{\odot}$ models. The vertical axis depicts the mass 
ratios of a model where the $^{12}$C$\,$+$^{16}$O reaction is fully suppressed, denoted by ``off'', and a model
where the rate given by \citet{Ca88} is considered, denoted by ``on''. The intermediate-mass elements show significant sensitivity to this rate, unlike 
the Fe-peak elements. The individual points are colored based on their mass abundances when the reaction is not included.}
\label{moffmon}
\end{figure}

Though there are likely additional differences due to the methods used in
these computations, we identify the $^{12}$C$\,$+$^{16}$O reaction 
rate as a significant source of the spread seen in Figure \ref{Comparison}. A precise
determination of the cross-section for this reaction remains elusive.
This is largely because the cross-section at stellar energies is in a
non-resonance region, where the cross-section is determined by the
interference between several broad resonances. In addition, high
energy tails of subthreshold levels whose properties are challenging
to determine directly can also complicate the extrapolation of the data into the Gamow range
\citep[e.g.,][]{Bu15,Fa17}. At temperatures $\simeq$ 4$\times$10$^{9}$ K, 
where the $^{12}$C$\,$+$^{16}$O rate is most influent, the Gamow peak 
of this reaction is 7.7$\,\pm\,$1.9 MeV. This rate impacts the Ca/S yield because of its
relation with the abundance of alpha particles, $\rm{M_{Ca}/M_{S}}
\propto X_\alpha^2$.

Given these theoretical uncertainties, the reaction was not included
in the model grid from \cite{Ya15} shown in Figures
\ref{CrFeCaS} and \ref{CaSeta}. However, the results 
from that paper, which are based on the
Fe-peak elements ($\rm{M_{Ni}/M_{Fe}}$, $\rm{M_{Mn}/M_{Fe}}$), remain
valid. To prove this, we run an additional DDTc, 5.4-Z$_{\odot}$ model
where the reaction is included (using the rate given by \citealt{Ca88}) and 
show the effect on the total mass yields in Figure \ref{moffmon}. 
The Fe-peak yields are insensitive to the $^{12}$C$\,$+$^{16}$O 
rate, but the Ca and S yields vary drastically. This could affect our 
inferred $Z_{eq}$ values given the small error bars in our 
measurements (see Section \ref{Observational}).

To study the effect of this reaction rate on the overall
$\rm{M_{Ca}/M_{S}}$ yield, we run additional \mch\ and
sub-\mch\ models. The \mch\ are calculated analogously to the ones in
Section \ref{Interpr_models}, although with an increased central
density $\rho_{\rm{c}} = 3 \times 10^{9} \, \rm{g \, cm^{-3} }$. The
sub-\mch\ are obtained with the methods used in \citet{Mi16} and 
described in \citet{To16}, applied in one dimension, and using the 
reaction networks provided by the Modules for Experiments in Stellar 
Astrophysics (\texttt{MESA}) for post-processing instead of 
\texttt{Torch} \citep{To16}. These two additional model grids 
give similar, though understandably not quite identical, yields for
the same $^{12}$C$\,$+$^{16}$O rate. We introduce several attenuation factors 
$\xi_{\rm{CO}}$: 0, 0.7 and 0.9 for the DDTs, and 0,
0.7, 0.9 and 0.99 for the sub-\mch. We use the rate given by
$\lambda = (1 - \xi_{\rm{CO}}) \, \lambda_{\rm{CF88}}$, 
where $\lambda_{\rm{CF88}}$ is the standard $^{12}$C$\,$+$^{16}$O rate \citep{Ca88},
so $\xi_{\rm{CO}} = 0, 1$ corresponds to null and full suppression, respectively.

\placefigure{f7}
\begin{figure*}
\centering
\includegraphics[scale=0.30]{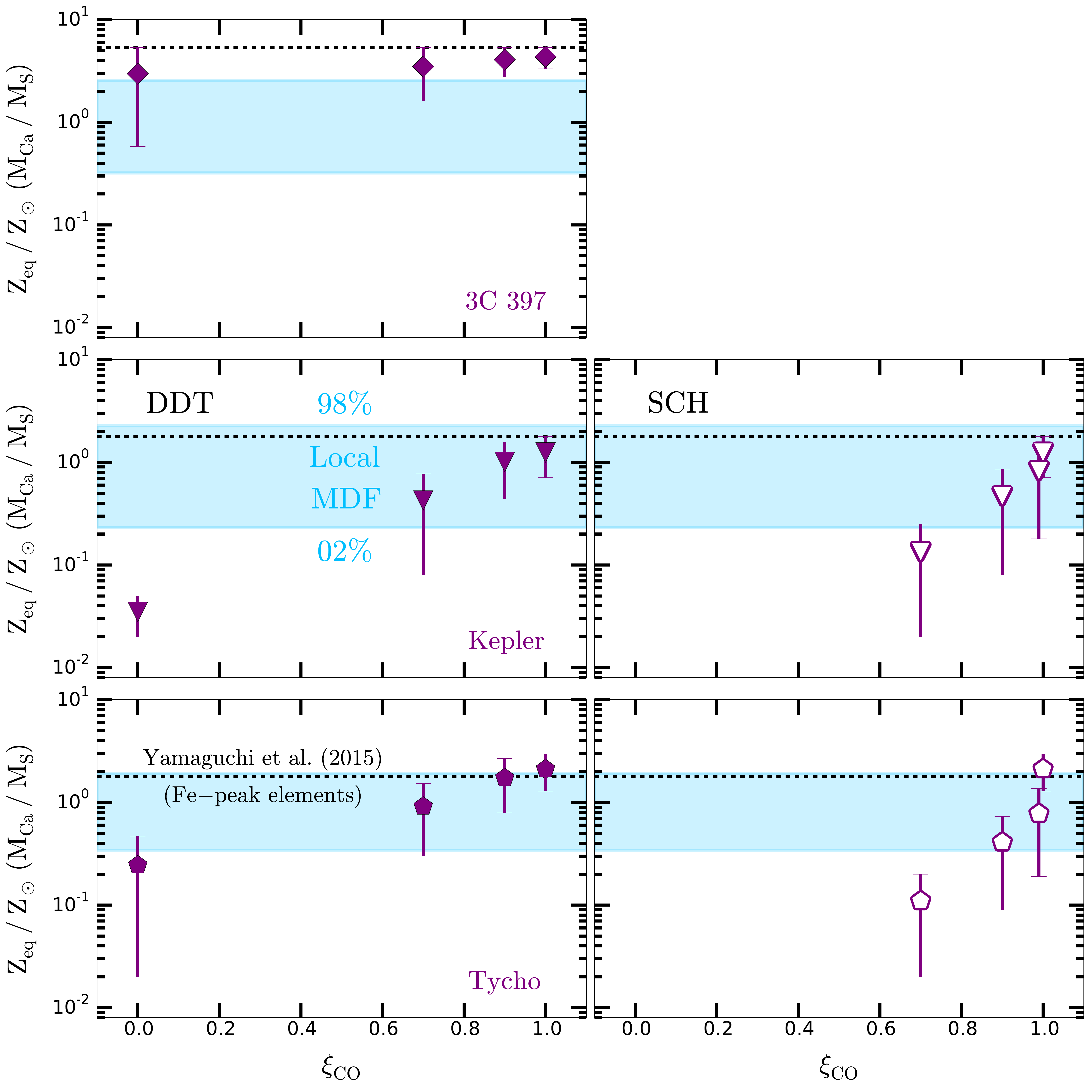}
\caption{Effect of different attenuations factors $\xi_{\rm{CO}}$ acting over the $^{12}$C$\,$+$^{16}$O reaction rate on the inferred equivalent metallicities $Z_{eq}$ for
3C 397, Kepler and Tycho. Values shown for $\xi_{\rm{CO}}=1$ are the same as those interpolated from Figure \ref{CrFeCaS} and displayed in Figure \ref{MDF}. The black,
dashed lines depict the equivalent metallicities found by \citet{Ya15}, whereas the blue, shaded regions represent the local MDF for the Milky Way disk in the
environment of each SNR (numbers indicate percentiles). Left: \mch-models. Right: sub-\mch\ models. We note that 3C 397 is not compatible with the latter. While the inferred neutron excess, 
$Z_{eq}$, is lower with the uncertain $^{12}$C$\,$+$^{16}$O reaction included ($\xi_{\rm CO}<1$), 3C 397 still shows evidence of an elevated metallicity compared to the other remnants.}
\label{xiCO}
\end{figure*}

Why does varying the $^{12}$C$\,$+$^{16}$O rate change the Ca/S mass ratio?
First, consider the case where the $^{12}$C$\,$+$^{16}$O rate is zero.  At
oxygen burning temperatures, oxygen could react with itself to mainly
produce $^{16}$O($^{16}$O,$\alpha$)$^{28}$Si. We will refer to this as
the ``alpha-poor'' branch since only one $\alpha$-particle is
produced.  Alternatively, oxygen can photodisintegrate to produce
carbon, 2($^{16}$O)$\,$+$\,$2$\gamma$ $\rightarrow$ 2($^{12}$C)$\,$+$\,$2$\alpha$.
If this carbon only recombines with the emitted $\alpha$-particle to
produce $^{16}$O, then this equilibrium loop is not interesting for
our purposes here. The other option is for carbon to burn with itself
to mainly produce $^{12}$C($^{12}$C,$\alpha$)$^{20}$Ne.
Photodisintegration of $^{20}$Ne then returns the nuclear flows to
$^{16}$O via $^{20}$Ne($\gamma$,$\alpha$)$^{16}$O. The net flow of this
oxygen cycle is $^{16}$O($^{16}$O,4$\alpha$)$^{16}$O.  We will refer to
this as the ``alpha-rich'' branch since four $\alpha$-particles are
produced.  The alpha-poor and alpha-rich branches compete with each
other.  Which branch dominates depends on the thermodynamic conditions
and reaction rates.  If the alpha-poor branch wins, then the $^{32}$S
and $^{40}$Ca abundances will be low.  If the alpha-rich branch wins,
then the $^{32}$S and $^{40}$Ca abundances will be high.
 
Now consider the case where the $^{12}$C$\,$+$^{16}$O rate is nonzero.
The $^{12}$C$\,$+$^{16}$O branching ratios are not important because the main
products from this reaction ($^{24}$Mg $^{27}$Al, and $^{27}$Si) ultimately
produce $^{28}$Si. That is, the net nuclear flow is $^{12}$C$\,$+$^{16}$O
$\rightarrow$ $^{28}$Si.  If $^{12}$C burns only by reactions with
$^{16}$O, the reaction flow for oxygen photodisintegration is
$^{16}$O$\,$+$\,\gamma$ $\rightarrow$ $^{12}$C$\,$+$\,\alpha$, then 
$^{12}$C$\,$+$^{16}$O $\rightarrow$ $^{28}$Si.
This is the same single $\alpha$-particle yield as the alpha-poor
branch \citep{Wo71}.  Thus, the net effect of a nonzero $^{12}$C$\,$+$^{16}$O rate is
to assist the alpha-poor branch, to produce less $\alpha$-particles.

\citet{De14} and \citet{Mi16} showed that the Ca/S mass ratio in SN Ia ejecta
scales as the square of the $\alpha$-particle abundance. Increasing
the $^{12}$C$\,$+$^{16}$O rate (decreasing $\xi_{\rm{CO}}$) suppresses the
$\alpha$-particle abundance, which in turn decreases $\rm{M_{Ca}/M_{S}}$.
There is less sensitivity to the $^{12}$C$\,$+$^{16}$O rate at higher
metallicity (more $^{22}$Ne) because the increased neutron richness
opens additional channels for $\alpha$-particles, so that the
action of the $^{12}$C$\,$+$^{16}$O reaction to shift $\alpha$-particle
flows toward the alpha-poor branch is less important.

In Figure \ref{xiCO}, we show how the various $^{12}$C$\,$+$^{16}$O rate
multipliers affect the determination of $Z_{eq}$ for 3C 397, Kepler
and Tycho by analyzing $\rm{M_{Ca}/M_{S}}$ vs $\rm{M_{Cr}/M_{Fe}}$
 and linearly interpolating within the model grids as done for Figure \ref{MDF}.
We choose these remnants because our inferred $Z_{eq}$ estimates agree
with previous measurements based on Fe-peak nuclei, which are not affected
	by the $^{12}$C$\,$+$^{16}$O rate (see the discussion in Section
\ref{Interpr_models}). In order to recover $Z_{eq}$
values that are consistent with the ones found by \cite{Ya15}, the suppression factor has to be 
at least of the order of $\xi_{\rm{CO}} = 0.9$ (attenuation $\gtrsim \,$90\%). 
We conclude that the $^{12}$C$\,$+$^{16}$O rate is attenuated in nature, but 
we emphasize that a more in-depth analysis is required to get to the bottom 
of this newly identified problem in SN Ia nucleosynthesis. For the purposes 
of this work, we point out that the correspondence between
$\rm{M_{Ca}/M_{S}}$ values and equivalent progenitor metallicities in our
Figures \ref{CrFeCaS}, \ref{CaSeta} and \ref{MDF} is tentative and might need to
be revised in the future. This certainly complicates our analysis, but
it does not invalidate our main conclusions that (1) the
neutronization in SN Ia ejecta appears to be high, given the values of
$\rm{M_{Ca}/M_{S}}$ measured in G337.2$-$0.7 and 3C 397 and the
dependence between $\rm{M_{Ca}/M_{S}}$ and neutronization identified
by \cite{De14}, and (2) because the Ca/S mass ratio in SNR N103B in
the LMC is comparable to that of Milky-Way-type Ia SNRs, it seems unlikely
that progenitor metallicity alone can be responsible for this high
neutronization.

\section{Conclusions} \label{Conclusions}

We have inferred the neutronization in the ejecta of five Type Ia SNRs (3C 397, N103B, G337.2$-$0.7, Kepler and Tycho) from their
X-ray spectra, using a new method based on the sensitivity of the Ca/S yield to $\eta$ discussed in \cite{De14}. The neutronization
inferred for N103B, in the LMC, is comparable to the values determined for Tycho and Kepler, in the Milky Way, which indicates that progenitor 
metallicity cannot be the only source of neutrons in SN Ia ejecta.

By comparing to a grid of SN Ia explosion models, we have translated our measured Ca/S mass ratios to equivalent progenitor metallicities,
which can be compared to the MDFs in the Milky Way and the LMC. These comparisons rule out progenitor metallicity as 
the sole source of neutrons for 3C 397, G337.2$-$0.7, and N103B. This represents a conundrum for SN Ia progenitors. Since our measurements 
are not affected by n-NSE and progenitor metallicity is discarded, the only possible source of neutronization left that we know can affect
 the whole ejecta is C simmering. Recent models of simmering by \cite{MR16} indicate that the highest level of neutronization is ${\simeq} \, 0.2 \, Z_{\odot}$,
  which is too low to explain the observations. This implies that either there is a fourth, as yet unidentified, source of neutronization in SN Ia progenitors, 
or that these simmering models do not capture the full phenomenology of C simmering. Lately, \cite{Pier17} have suggested that the simmering contribution to
$\eta$ is higher than that of \cite{MR16}, but more work is needed to understand the differences between both analyses.

We have also identified an issue affecting most SN Ia nucleosynthesis calculations in the literature. The Ca/S mass ratio in the final yields is very 
sensitive to the precise value of the $^{12}$C$\,$+$^{16}$O reaction rate (see Figure \ref{moffmon}), with the most widely used rate 
value leading to Ca/S mass ratios that are too low to reproduce our measurements by a factor of ${\sim}\ $2 (shown in Figure \ref{Comparison}). Given the 
excellent correspondence between the SN Ia model grid used in this work, where this reaction rate is not included, and our $\rm{M_{Ca}/M_{S}}$ measurements, 
we conclude that the  $^{12}$C$\,$+$^{16}$O reaction rate must be suppressed in nature by a potentially large factor. A preliminary exploration 
of SN Ia nucleosynthesis calculations with varying degrees of suppression in the  $^{12}$C$\,$+$^{16}$O reaction, displayed in Figure \ref{xiCO}, 
confirms this conclusion, but a more detailed analysis is needed to get to the bottom of this issue \citep[e.g.,][]{Bu15,Fa17}. Until this study is completed, our estimates 
of $Z_{eq}$ must be considered tentative, and will need to be revised.

We emphasize that our main results are not sensitive to these details. The values of $\rm{M_{Ca}/M_{S}}$ measured in our SNRs G337.2$-$0.7 and 3C 397 
do require a high degree of neutronization in SN Ia ejecta, by virtue of the effect discovered by \cite{De14}. Most importantly, the fact that SNR N103B 
in the LMC shows a Ca/S mass ratio similar to those of Milky Way SNRs like Tycho strongly suggests that metallicity alone cannot be the origin of this high 
neutronization. Unless a new source of neutrons in SNe Ia is identified, the simplest explanation for this high neutronization is that a large fraction of 
SNe Ia in the local universe explode close to \mch\, after developing a large convective core through carbon simmering.

\acknowledgments We are grateful to Michael Hayden for sharing his APOGEE data with which we generated the MDF for the Milky Way disk. We also thank Luc 
D\'{e}ssart for sharing his supernova models with us, Peter H\"{o}flich for insightful discussions about the nucleosynthesis in these models, and
Stan Woosley for his help to interpret his nucleosynthetic yields. H.M.-R., C.B., H.Y. and S.P. are funded by the NASA ADAP grant NNX15AM03G S01. 
H.M.-R. also acknowledges support from a PITT PACC Fellowship, and C.B. from the OCIW Distinguished Visitor program at the Carnegie Observatories. 
E.B. is supported by the MINECO-FEDER grant AYA2015-63588-P. This research has made use of NASA's Astrophysics Data System 
(ADS, \url{http://adswww.harvard.edu/}) and of the HEASARC spectral data base (\url{http://heasarc.gsfc.nasa.gov/}).

\software{\texttt{FTOOLS} \citep{Bl95}, \texttt{XSPEC} \citep{Ar96}, \texttt{Flash} \citep[\url{http://flash.uchicago.edu/site},][]{Fr00,Du12}, \texttt{SAOIMAGE DS9} \citep{Jo03}, \texttt{Matplotlib} \citep{Hun07}, \texttt{IPython} \citep{PeG07}, \texttt{MESA} \citep[\url{http://mesa.sourceforge.net/},][]{Pa11,Pa13,Pa15}, \texttt{Numpy} \citep{vaW11}, \texttt{AtomDB} \citep{Fo12, Fo14}, \texttt{Astropy} \citep{Astro13}, \texttt{SN Ia Flash modules} \citep[\url{http://pages.astronomy.ua.edu/townsley/code/},][]{To16}, \texttt{Python} (\url{https://www.python.org/}).}

\vspace{0.5 cm}


\twocolumngrid

\bibliography{Paper_N103B}

\end{document}